\documentclass[9pt,twocolumn,twoside]{opticajnl}
\journal{opticajournal} 
\usepackage{dblfloatfix}

\setboolean{shortarticle}{false}

\usepackage{lineno}

\title{Designing Noise-Robust Quantum Networks Coexisting in the Classical Fiber Infrastructure }

\author[1,*]{Jordan M. Thomas}
\author[1,3]{Gregory S. Kanter}
\author[1,2]{Prem Kumar}

\affil[1]{Center for Photonic Communication and Computing, Department of Electrical Engineering, Northwestern University, 2145 N. Sheridan Rd., Evanston, IL. 60208-3112, USA}
\affil[2]{Department of Physics and Astronomy, Northwestern University, 2145 Sheridan Road,
Evanston, IL 60208-3112, USA}
\affil[3]{NuCrypt, LLC, 1460 Renaissance Dr \#205, Park Ridge, IL 60068}
\affil[*]{jordanthomas2025@u.northwestern.edu}

\begin{abstract}
The scalability of quantum networking will benefit from quantum and classical communications coexisting in shared fibers, the main challenge being spontaneous Raman scattering noise. We investigate the coexistence of multi-channel O-band quantum and C-band classical communications. We characterize multiple narrowband entangled photon pair channels across 1282 nm-1318 nm co-propagating over 48 km installed standard fiber with record C-band power (>18 dBm) and demonstrate that some quantum-classical wavelength combinations significantly outperform others. We analyze the Raman noise spectrum, optimal wavelength engineering, multi-photon pair emission in entangled photon-classical coexistence, and evaluate the implications for future quantum applications.
\end{abstract}

\setboolean{displaycopyright}{false} 

\begin{document}

\maketitle

\section{Introduction}

\begin{figure*}[b]
\centering
\includegraphics[width=\textwidth]{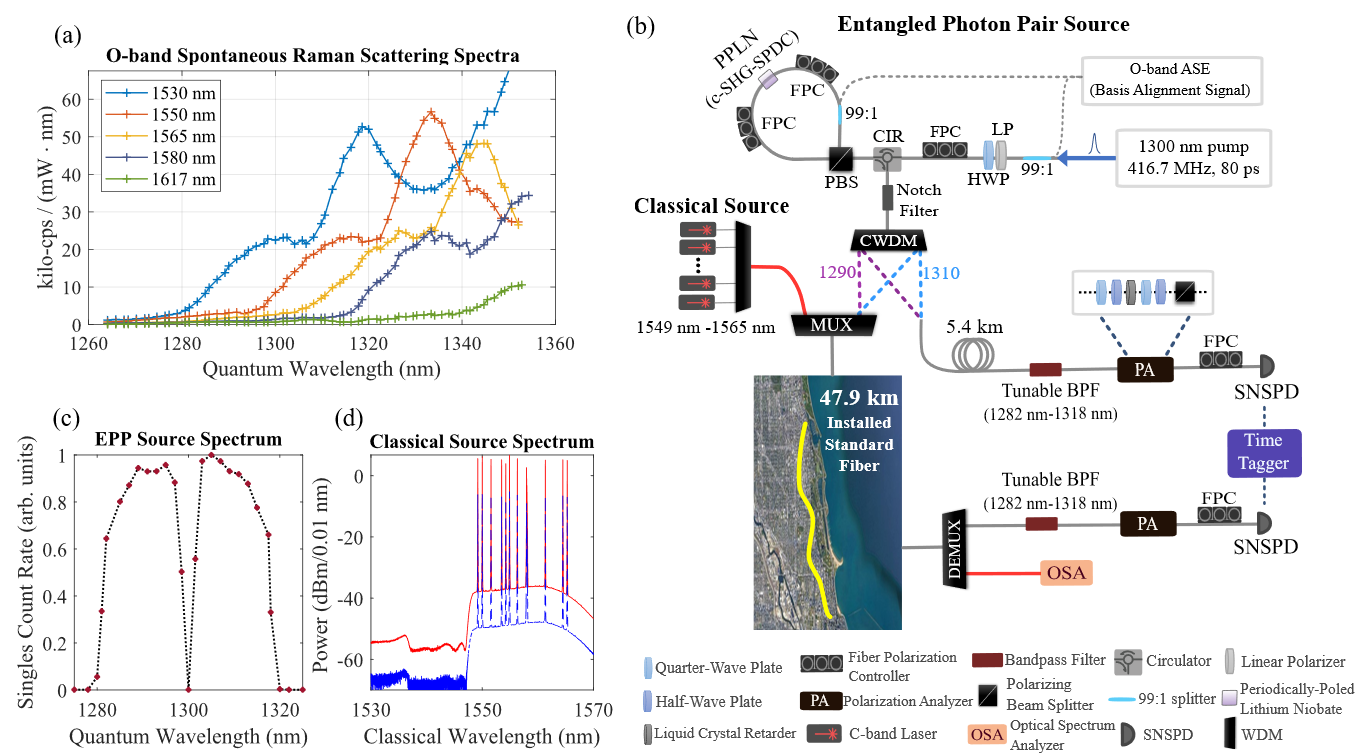}
\caption{(a) Spontaneous Raman scattering spectrum across the O-band from C- and L-band classical pumps. (b) Experimental design for polarization entangled photon-classical coexistence over 47.9 km of installed standard fiber. A classical WDM C-band source is multiplexed in with either <1300 nm or >1300 nm entangled photons to co-propagate over the installed fiber, whereas the other is routed 5.4 km spooled dark fiber. Narrowband tunable wavelength filters at each node characterize the noise resilience of various signal-idler channel pairs spanning 1282 nm-1318 nm. (c) Polarization entangled photon pair (EPP) source’s transmitted joint spectrum. Entangled photons <1300 nm are more robust to spRS noise from classical light >1550 nm compared to photons >1300 nm based upon (a). (d) Classical WDM C-band source input (red)-output (blue) spectrum.}
\label{fig:fig1}
\end{figure*}

Recent years have seen a surge of research interest towards realizing a deployed quantum internet (QI) 
\cite{kimble_quantum_2008,wehner_quantum_2018}, fueled by the discovery of applications beyond the realm of classical physics such as physically secure communications via quantum key distribution (QKD) \cite{ekert_ultimate_2014}, quantum enhanced measurements \cite{komar_quantum_2014, PhysRevLett.120.080501, PhysRevLett.109.070503}, improved scalability \cite{monroe_scaling_2013} and secure use \cite{broadbent_universal_2009} of quantum computers, amongst others. As a result, there is a global initiative for realizing the QI precursor, namely quantum networks, beyond laboratory settings. Achieving this goal requires in part the ability to distribute high-fidelity quantum entangled states between nodes, where any extraneous noise can severely limit quantum operations. \\ \indent An accessible approach for deploying quantum networks is to leverage the optical fiber infrastructure and telecommunications technology to generate and route telecom band quantum states between quantum nodes. Since both quantum and classical communications have interest in operating in installed fibers and the fact that most are already populated with classical communications, the scalability of quantum networking depends on quantum and classical signals “coexisting” in the same fibers as this would increase the number of available fiber routes and reduce costs of leasing and maintaining fibers. \\ \indent Although quantum-classical coexistence has been investigated since the early days of fiber-based quantum communications \cite{townsend_simultaneous_1997}, most studies have focused on single photon or continuous variable QKD \cite{townsend_simultaneous_1997, chapuran_optical_2009, Aleksic:15, dynes_ultra-high_2016,  kumar_coexistence_2015, eraerds_quantum_2010, qi_feasibility_2010, patel_coexistence_2012, 
PhysRevA.95.012301, Geng:21, Geng:23,gavignet_co-propagation_2023, Mao:18}. However, with the rapid progress towards deploying multi-node entangled photon pair-based networks (e.g., \cite{koduru_trusted_2020-1, alshowkan_reconfigurable_2021, clark_entanglement_2023, chung_design_2022}), quantum teleportation \cite{valivarthi_quantum_2016}, entanglement swapping \cite{sun_entanglement_2017-2}, and entangled quantum memories \cite{yu_entanglement_2020, du_elementary_2021} in installed dark fibers, it is timely to begin considering the unique design and control methods needed for achieving high-fidelity advanced quantum network capabilities alongside high data rate classical communications in shared fibers. In this regard, there have been relatively fewer investigations of entangled photon pair distribution \cite{1637099,ciurana_entanglement_2015,sauge_narrowband_2007,10.1117/12.2230222,Yuan:19, Thomas:22} or Bell state measurements (BSMs) \cite{valivarthi_measurement-device-independent_2019, berrevoets_deployed_2022} coexisting with independent classical communications, with only a few demonstrated over installed fiber \cite{Thomas:22,berrevoets_deployed_2022}. These are key building blocks for future quantum networks, meaning the design of noise-robust entangled photon sources, optimal network routing, measurement systems, etc., is a crucial task towards scaling more advanced quantum networks beyond dark fibers. \\ \indent The main challenge for quantum-classical integration in standard fibers is the generation of noise photons via spontaneous Raman scattering (spRS) of high-power classical light. Due to the strength of Raman gain at small frequency offsets, placing both quantum and classical signals in the same telecom band significantly constrains classical power levels (and thus classical data rates) before quantum signals are entirely obscured by noise. Since most classical communications operate in the C-band (1530 nm-1565 nm) or L-band (1565 nm-1625 nm), this significantly limits the ability for quantum signals to operate at these wavelengths, albeit being preferred in dark fibers due to low absorption loss. For example, previous C-band entangled photon \cite{1637099,Yuan:19} or BSM applications \cite{valivarthi_measurement-device-independent_2019} have required attenuating coexisting C-band powers to below -10 dBm, which severely restricts the data capacity of the fiber. One approach to remedy this issue is to offset the quantum wavelength to be significantly shorter than the classical wavelength such that spRS noise is greatly suppressed by lower Raman gain and anti-Stokes detuning. However, the wavelength detuning should not be so large as to exit the low-loss single-mode propagation regime of standard fiber. In practice this has corresponded to moving quantum signals to the O-band (1260 nm-1360 nm) \cite{townsend_simultaneous_1997}, where 1310 nm is most common. While the O-band has somewhat higher loss compared to the C-band (a difference of $\sim$ 0.1 dB/km), the lower noise in the O-band can allow for higher QKD rates in fibers where C-band spRS noise dominates \cite{PhysRevA.95.012301}. This approach has allowed recent entangled photon pair experiments to reach mW-levels of coexisting 1550 nm power \cite{Thomas:22}. However, a few 1310 nm single-photon QKD systems have coexisted with 10’s of mW (>10 dBm) aggregate powers of wavelength division multiplexed (WDM) C-band classical communications encoding as high as many terabits per second (Tb/s) rates over 50 km to 80 km spooled fiber \cite{PhysRevA.95.012301, Geng:21, Geng:23,gavignet_co-propagation_2023}, the most accomplished being coexistence with 21 dBm C-band power over 66 km installed commercial backbone fiber \cite{ Mao:18}. \\ \indent To our knowledge, the distribution of entangled photons in such extreme power-fiber length scales has yet to be demonstrated. Further, to our knowledge the ability to simultaneously transmit multiple O-band quantum channels, and whether some outperform others, have not been evaluated in these regimes. Like classical networking, WDM technology can greatly enhance quantum network capabilities. Some examples include spectrally carving broadband entangled photon pairs to network many user pairs from a single central node \cite{koduru_trusted_2020-1, alshowkan_reconfigurable_2021, clark_entanglement_2023}, multiplexing quantum channels into a single fiber for inter-connecting distant local area networks, and spectrally multiplexed repeaters \cite{sinclair_spectral_2014}, amongst others. Thus, having the ability to transmit many O-band WDM channels of high-fidelity entangled photons alongside high WDM C-band powers in installed fiber would be a significant step towards deploying advanced quantum network applications beyond dark fibers. \\ \indent In this paper, we show that high-fidelity O-band entangled photon WDM quantum network applications can coexist with at least 18 dBm aggregate WDM C-band power over 47.9 km of installed standard fiber. We verify this ability with multiple 50 GHz filtered entangled photon pair channels spanning the 1282 nm-1318 nm band, demonstrating for the first time not only the feasibility of entangled photon pairs coexisting alongside Tb/s C-band classical communications in installed fiber, but also that many O-band quantum channels can simultaneously coexist for quantum applications. \\ \indent However, we find some channels are more robust to spRS noise than others. We first characterize the shape of the O-band spRS spectrum from various C- to L-band classical pumps to evaluate the least noisy O-band channels for a given classical source. These measurements reveal that quantum-classical systems can be universally improved over the 1310 nm-1550 nm standard, where spRS is reduced roughly an order of magnitude by changing the quantum signals by 10’s of nm’s to the shorter O-band wavelengths or using L-band classical communications. \\ \indent Based on these results, we focus on designing noise robust entanglement sources, quantum network routing, and measurement systems. We design a broadband O-band polarization entangled photon pair source with a ~40 nm wide joint-spectrum centered around 1300 nm to demonstrate that entangled photons at wavelengths <1300 nm are more robust to spRS compared to >1300 nm from an 11-channel 1549 nm-1565 nm C-band WDM classical source due to the shape of the spRS spectrum across the O-band channels.\\ \indent We compare routing either the <1300 nm or >1300 nm entangled photons in a 2x2 switch configuration to co-propagate with as high as 18.1 dBm aggregate WDM C-band launch power over 47.9 km installed fiber, whilst the other photon is sent over 5.4 km spooled dark fiber. Characterizing the 1282 nm-1318 nm band using tunable 50 GHz filters, we find that routing channels <1300 nm perform nearly identical to dark fiber, whereas channels >1300 nm suffer from higher spRS noise. We further evaluate how multi-photon pair emission in spontaneous parametric entangled photon pair sources influences entangled photon-classical coexistence and how this impacts quantum applications. We then perform quantum state tomography (QST) \cite{ALTEPETER2005105} on the optimal quantum channels, showing only a 0.4\%, 0.9\%, and 1.5\% reduction in fidelity compared to dark fiber with 14, 16.2 and 18.1 dBm C-band launch powers, respectively. Lastly, we discuss the implications for future advanced quantum network applications such as switch-based multi-node quantum networks, quantum teleportation, entanglement swapping, and beyond.

\section{Designing Spontaneous Raman Scattering Noise Robust Coexisting Quantum-Classical Networks}

\subsection{Optimal Choice for Quantum-Classical Wavelengths Dictated by the Shape of the spRS Spectrum}

\indent Although 1310 nm quantum signals are the most common choice to coexist with C-band classical communications (CCs), it is not obvious that this is generally the most noise robust quantum channel. The Raman gain spectrum is a function of the frequency offset between quantum-classical signals $\Omega = \nu_{\text{Cl}}- \nu_{\text{Qu}}$ with a broad bandwidth (BW) and complex shape. Within the spectrum there are many local maximum and minima due to the multiple vibrational modes of fused silica molecules in fiber \cite{Hollenbeck:02}, where slight changes in $\Omega$ can have significantly different Raman gain. The O-band spans 100 nm, whilst the C- to L-bands combined span 95 nm, resulting in a wide range of possible frequency offsets ($\Omega \sim $ 32 THz to 54 THz) where multiple modes can reside within the O-band. Therefore, some O-band channels can be more susceptible to spRS than others depending on the choice of quantum-classical channel(s). Beyond the C-band, L-band technology has been developed in large part to enhance the available CC BW. As we will show, it can be a useful approach for significantly reducing spRS across the O-band whilst still allowing a broad BW for WDM CCs.

To compare the strength of spRS across the O-band from various C- or L-band classical sources, we measure the O-band spectrum of forward spRS from classical pumps at wavelengths $\lambda_{\mathrm{Cl}}$ = 1530 nm, 1550 nm, 1565 nm, 1580 nm, and 1617 nm with 2.05 dBm launch power over 25 km SMF-28 fiber. We measure spRS rates at O-band wavelengths $\lambda_{\mathrm{Qu}}$ using a 0.25 nm full width-half maximum (FWHM) filter and a superconducting nanowire single photon detector (SNSPD) with a 0.92 $+/-$ 0.03 efficiency across the O-band.

The results are shown in Fig. \ref{fig:fig1}(a). We see that spRS is highly dependent on the choice of quantum and/or classical wavelengths. O-band channels $<1290 \mathrm{~nm}$ have roughly an order of magnitude less noise compared to $1310 \mathrm{~nm}$ from $\lambda_{\mathrm{Cl}}$ $=1550 \mathrm{~nm}$, and $1330 \mathrm{~nm}$ has roughly twice as much as 1310 $\mathrm{nm}$. If we instead vary the classical source, $\lambda_{\mathrm{Qu}}=1310 \mathrm{~nm}$ has an analogous noise improvement with $\lambda_{\mathrm{Cl}}>1580 \mathrm{~nm}$ compared to $1550 \mathrm{~nm}$, and $\lambda_{\mathrm{Cl}} \sim 1617 \mathrm{~nm}$ obtains this improvement up to $\lambda_{\text {Qu }}<1340 \mathrm{~nm}$. Since longer O-band wavelengths have lower loss, O-band/L-band coexistence is an appealing approach for noise-robust quantum-classical networking. However, interestingly we show in Section 2.\ref{2b} that using higher loss channels does not generally correspond to lower quantum rates for a particular error rate given a sufficient noise improvement over lower loss channels.

We note that measurements of the spRS spectrum from a $1550 \mathrm{~nm}$ pump in \cite{Aleksic:15} were fit to a 13 vibrational mode model of the Raman gain spectrum \cite{Hollenbeck:02} down to $1250 \mathrm{~nm}$ offsets $(\Omega \sim 47 \ \mathrm{THz})$. We observe a weak 14th mode centered at $\Omega \sim 48 \ \mathrm{THz}$. The implications for wavelength engineering are that there is a relatively flat Raman gain between $39 \ \mathrm{THz}$ to $47 \ \mathrm{THz}$, wherein spRS rates decrease by $\sim 3$ times across this band due to a decreasing phonon population factor before it increases slightly from the additional mode, then decreases again at $\Omega>50 \ \mathrm{THz}$. On one hand it allows a very broad $\mathrm{BW}$ with low noise, however it does restrict the ability to dramatically reduce noise by increasing $\Omega$. Only above 50 THz separations does spRS significantly decrease further. Although this restricts both systems, it could find applicability in situations extremely susceptible to noise.

All the above considerations can be applied universally in quantum-classical engineering. To coexist with the C-band, systems currently operating at $1310 \mathrm{~nm}$ could use lower O-band $\lambda_{\text {Qu}}$ or explore moving CCs to the L-band for reducing noise at $1310 \mathrm{~nm}$. In terms of designing broadband EPP sources, centering the joint-spectrum within the lower noise band would improve both signal-idler channels, or one could selectively route \cite{alshowkan_reconfigurable_2021, chung_design_2022} specific channels that are most robust to spRS given the spectrum across the quantum BW. Recently, quantum memories have been entangled via the transmission of $1342 \mathrm{~nm}$ \cite{yu_entanglement_2020} and $1324 \mathrm{~nm}$ \cite{du_elementary_2021} photons in dark fiber. These wavelengths generally will experience more noise from coexisting C-band light than sub-1310 nm wavelengths but do have slightly lower propagation loss. If these systems are to exist beyond dark fibers, using $\sim 1610 \mathrm{~nm}$ $\mathrm{CCs}$ or frequency converting to shorter O-band $\lambda_{\text {Qu}}$ can reduce spRS noise $>30$ times compared to some C-band channels.

\subsection{Designing Robust O-band Entanglement Sources and Receivers Coexisting with C-, L-band Classical Communications}

We now experimentally demonstrate how the properties of spRS can be exploited to design noise-robust coexisting quantum-classical networks, where we focus on O-band entangled photon pair (EPP)-based quantum network (QN) applications coexisting alongside high power WDM C-band CCs in installed fiber. 

The experimental design is shown in  Fig. \ref{fig:fig1}(b). We generate polarization EPPs via cascaded second harmonic generation-spontaneous parametric down conversion (c-SHG-SPDC) \cite{Arahira:11} in a single fiber-coupled periodically poled lithium niobate waveguide that is phase matched for SHG at 1300 nm. A 1300 nm continuous wave (CW) laser is intensity modulated to create a stream of 80 ps FWHM pulses with a 416.7 MHz repetition rate. The pump is split 50:50 in a polarization Sagnac loop to generate the quantum state $|\Psi\rangle \propto$ $1/ \sqrt{2}\left(|\mathrm{HH}\rangle+\mathrm{e}^{\mathrm{i} \varphi}|\mathrm{VV}\rangle\right) \otimes \iint \mathrm{d} \omega_{\mathrm{s}} \mathrm{d} \omega_{\mathrm{i}} F\left(\omega_{\mathrm{s}}, \omega_{\mathrm{i}}\right)\left|\omega_{\mathrm{s}}, \omega_{\mathrm{i}}\right\rangle$ where the signal-idler joint spectrum $F\left(\omega_{\mathrm{s}}, \omega_{\mathrm{i}}\right)$ is centered at 1300 nm and has a $\sim$ 40 nm FWHM. We then use notch filters to reject the pump followed by a 1290 nm-1310 nm coarse-WDM (CWDM) to separate signal and idler spectra, respectively. The transmitted signal-idler single photon spectrum is shown in  Fig. \ref{fig:fig1}(c). To compare routing either the 1290 or 1310 CWDM, we manually 2x2 switch each output to transmit either 47.9 km of installed underground fiber or 5.4 km of dark spooled fiber, wherein the 47.9 km is preceded by WDMs to combine O-band and C-band light for co-propagation.  We further include a built-in polarization basis alignment signal \cite{chung_design_2022,align} to allow for fast compensation of random birefringent rotations over the fiber links and align the $\left|\Phi^{+}\right\rangle$ Bell state between nodes. \\ \indent At quantum receiver nodes, we cascade WDMs to demultiplex quantum-classical channels and then isolate C-band light from the quantum passbands. Fundamentally, spectral-temporal filtering around the minimum number of a quantum signal’s modes minimizes the impact of multi-mode uncorrelated background noise \cite{bouchard_achieving_2021}. To narrowly filter in the time-frequency domains, we use 0.25 nm ($\sim$ 50 GHz) FWHM filters and 600 ps coincidence windows. The time-frequency filtering was limited by electronic and fiber jitter and the tunable filters used, where single mode detection could be achieved with some experimental modifications (see Section 3.\ref{3a}.) To characterize polarization entanglement, polarization analyzers at each node receive the source’s basis alignment signal to compensate fiber birefringence and perform basis projective coincidence detection. Each node uses low dark count ($\sim$ 100 counts/s) and high efficiency (0.92 +/- 0.03 across the O-band) Quantum Opus SNSPDs. Although our system benefits from high efficiency, the SNSPDs are free-running and thus detect noise outside of quantum pulse arrival times, where gating detectors could reduce the impact of spRS. \\ \indent A key point of this investigation is to evaluate the ability for both quantum O-band and classical C-band WDM networking to coexist with low noise and broad BWs. To evaluate this, the 50 GHz filters at signal-idler quantum nodes are tunable to characterize the impact of spRS on many narrowband channel pairs across 1282 nm-1318 nm when 2x2 routing the <1300 nm or >1300 nm entangled photons to co-propagate with a multi-channel C-band WDM source over the 47.9 km installed fiber. The classical source consists of 11 multiplexed C-band CW lasers across 1549 nm-1565 nm (Fig. \ref{fig:fig1}(d)). We amplify the C-band source with an erbium doped fiber amplifier followed by filters to reject out-of-band amplified spontaneous emission (ASE). Note that we verified the received spRS rates scaled linearly with launch power over the entire launch-power range, indicating Brillouin scattering has negligible effect in these experiments. From Fig. \ref{fig:fig1}(a) we see that the <1300 nm entangled photons should have significantly lower noise compared to the >1300 nm photons, allowing us to compare the performance of channels with various spRS noise levels. \\ \indent The $47.9 \mathrm{~km}$ installed fiber link is part of a local area network in the broader Illinois Express Quantum Network \cite{chung_design_2022, 10.1117/12.2588007}. We measure $12.7 \mathrm{~dB}$ net loss of the C-band WDM source, $20.9 \mathrm{~dB}$ at 1290 $\mathrm{nm}$, and $19.7 \mathrm{~dB}$ at $1310 \mathrm{~nm}$. The higher losses compared to ideal fiber is due to imperfect splicing and connections. As typical fiber has $\sim 0.33 \mathrm{~dB} / \mathrm{km}$ loss at $1310 \mathrm{~nm}$, these losses convert to distances of $\sim 60 \mathrm{~km}$ if ideal fiber was used. As spRS scales linearly with average launch power, the ability for quantum-classical systems to coexist is best quantified by the classical power levels and fiber lengths as opposed to a particular data modulation format or rates, where the possible data rates can be inferred from launched-received powers. Given the power in our WDM source is relatively evenly spread across the used BW, the spRS impact from our source versus if the BW were packed with WDM channels would not significantly differ. To compare a few power levels, we set launch powers such that the received aggregate power is either $\mathrm{P}_{\mathrm{Rx}}=1.3, \ 3.5$, or $5.4 \ \mathrm{dBm}$, corresponding to launch powers of $\mathrm{P}_{\mathrm{Tx}}=14, \ 16.2$, or $18.1 \ \mathrm{dBm}$, respectively. Such high powers are compatible with $\mathrm{Tb} / \mathrm{s}$ CCs over the $47.9 \mathrm{~km}$ installed fiber when compared to the coexistence experiments in \cite{PhysRevA.95.012301, Geng:21, Geng:23,gavignet_co-propagation_2023,Mao:18}.

\begin{figure}[tp]
\centering
\includegraphics[width=0.44\textwidth]{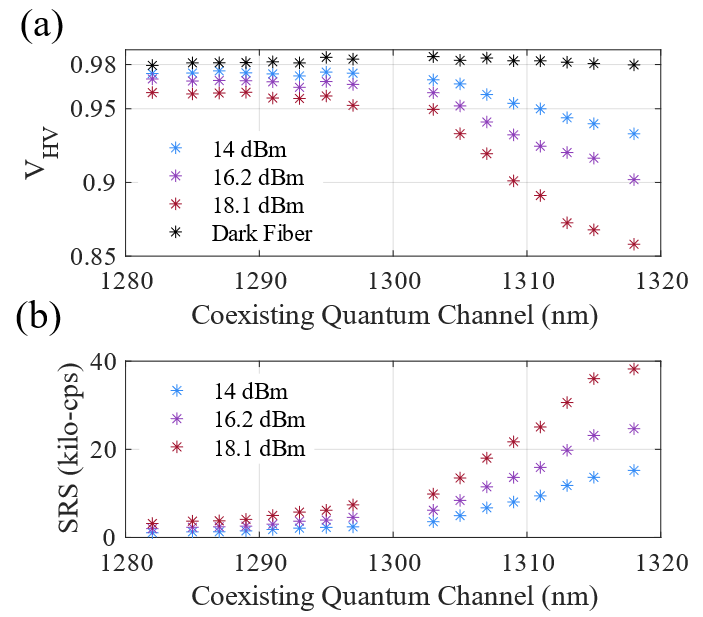}
\caption{(a) Two-photon interference visibility for 50 GHz filtered entangled photons between 1282 nm-1318 nm routed to co-propagate 47.9 km with 14 dBm, 16.2 dBm, or 18.1 dBm WDM C-band launch power. (b) spRS rates across each O-band channel from the WDM C-band source.}
\label{fig2}
\end{figure}

\section{O-band Entangled Photon Pair Quantum Networking in the Classical Fiber Infrastructure}

\subsection{Characterization of O-band WDM Entangled Photon Pair/C-band WDM Classical Coexistence}\label{2a}

We now characterize the performance of our O-band system to distribute multiple WDM channel pairs across the 1282 nm-1318 nm band whilst co-propagating over 47.9 km installed fiber with the 1549-1565 nm WDM source. \

To emulate an EPP node distributing multiple O-band channels in a QN, we fix the EPP source pump power (maintaining nearly fixed pair generation rates for all the channel pairs) and tune the signal-idler filters across the received joint-spectrum. We measure the two-photon interference (TPI) visibility $\text{V}_{HV}$ in the horizontal/vertical basis across multiple channel pairs when 2x2 routing the 1290 nm or 1310 nm CWDM spectra to co-propagate 47.9 km installed fiber with either 14 dBm, 16.2 dBm, or 18.1 dBm aggregate C-band launch power, where the other photon is routed 5.4 km spooled dark fiber. The EPP source is set to yield an average coincidence count rate (CCR) across all channel pairs of 44.3 coincidence counts per second (ccps) with a standard deviation of 6.1 ccps. Fig. \ref{fig2}(a) shows the measured $\text{V}_{HV}$ as we tune the 50 GHz filters across 1282 nm-1318 nm for each routing case and \ref{fig2}(b) shows the spRS rates in each channel due to the WDM C-band source. We also compare the measured visibility in dark fiber for reference. 

We find that the coexisting classical source has little impact (compared to dark fiber) on quantum channels $<1305$ $\mathrm{nm}$, but clearly reduces visibility for the $>1305 \mathrm{~nm}$ channels. Although $>1305 \mathrm{~nm}$ photons suffer, they still maintain visibilities $>85 \%$ up to $18.1 \mathrm{dBm}$ coexisting power, which is well above that required to violate Bell's inequality $(\mathrm{V}>70.7 \%$ \cite{clauser_proposed_1969}) and are therefore useful for quantum applications. However, the results clearly demonstrate the importance of optimized quantum-classical BW sources and routing strategies for designing robust and high-performance quantum-classical coexisting networks.

\subsection{Effect of Multi-Photon Pair Emission in Entangled Photon-Classical Coexistence}\label{2b}

In the previous section we operated at a fixed EPP source pump power to characterize each channel. However, the state fidelity in spontaneous parametric EPP sources highly depends on the photon pair generation probability due to multi-photon pair emission, and is dependent on signal-idler losses when dark count rates are non-negligible \cite{takesue_effects_2010}. Thus, the visibility for a given CCR is dependent on multi-photon pair emission, losses, as well as the spRS rates. Since spRS has a random time distribution from independent coexisting classical channels, the effect of spRS on EPP distribution should resemble a rise in dark count rates in the coexisting quantum channel(s). To evaluate this effect, we measure $\text{V}_{HV}$ as a function of increasing CCRs for dark fiber and each power setting. For simplicity, we fix the signal-idler filters at one channel pair (1287 nm, 1313 nm) corresponding to a pair of lower and higher noise channels found in Fig. \ref{fig2}. Background count rates from spRS scale linearly with launch power as 474.2 cps/mW and 62.1 cps/mW with installed fiber losses of 19.5 dB and 21.1 dB for 1313 nm and 1287 nm, respectively.

\begin{figure}[tp]
\centering
\includegraphics[width=0.45\textwidth]{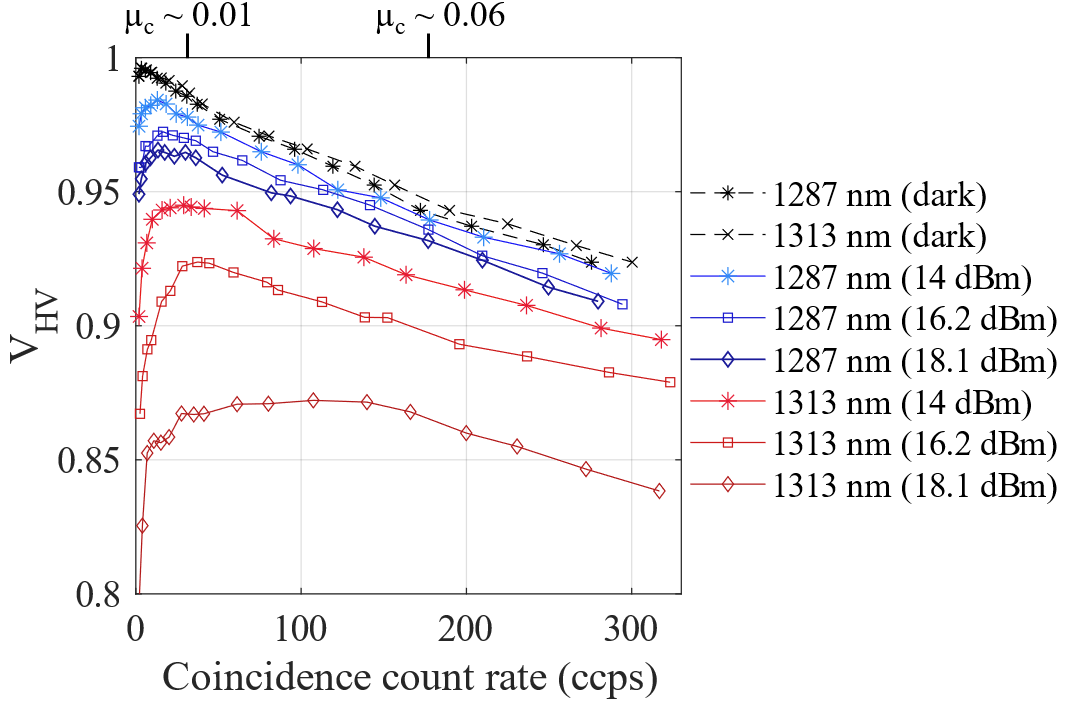}
\caption{Two-photon interference visibility as a function of coincidence count rates for either 1313 nm or 1287 nm entangled photons routed to co-propagate with 14 dBm, 16.2 dBm or 18.1 dBm aggregate C-band launch power over the 47.9 km installed fiber.}
\label{fig3}
\end{figure}

The results are shown in Fig. \ref{fig3}. We find that the dependence on CCR for the 1287 nm is only slightly impacted compared to dark fiber, whereas the 1313 nm channel is significantly more impacted. Interestingly, even though 1287 nm has higher loss, it always maintains a higher $\text{V}_{HV}$ compared to 1313 nm for a given CCR. Thus, we observe that sufficient noise improvements can allow lower O-band wavelengths, albeit having higher loss, to have the same CCRs as the lower loss channels and with higher visibility for a given rate. However, this may not be general across all systems, where all parameters such as signal-idler losses, noise improvements, etc. should be considered. 

For QN operations requiring high single photon fidelity such as BSMs, spontaneous parametric EPP sources usually operate at low mean photon pairs generated per pulse $\mu_{\mathrm{c}}$ to reduce multi-photon pair emission. For example, the quantum teleportation experiments in \cite{PRXQuantum.1.020317} operated at $\mu_{\mathrm{c}}<$ 0.01 to achieve high fidelity. In Fig. \ref{fig3}, we estimate $\mu_{\mathrm{c}}$ based on the dark fiber visibility ( $\mu_{\mathrm{c}} \approx 1 / \mathrm{V}-1$ \cite{takesue_effects_2010}) to evaluate how spRS impacts the ability to operate in this regime. We find that $1287 \mathrm{~nm}$ can allow high visibility near $\mu_{\mathrm{c}}<0.01$ due to the low noise. However, $1313 \mathrm{~nm}$ requires increasing $\mu_{\mathrm{c}}$ to overcome the much higher spRS noise, where $\mu_{\mathrm{c}}<0.01$ appears to be where $1313 \mathrm{~nm}$ suffers the most in comparison to $1287 \mathrm{~nm}$. When $\mu_{\mathrm{c}}$ is much higher $\left(\mu_{\mathrm{c}}>0.06\right)$, multi-photon pairs become a dominant source of noise and reduces improvement, however EPP sources are rarely operated at these levels for the above reasons. Thus, the lower noise wavelengths allow high single photon fidelity at high powers where noisier channels suffer the most, making the choice of quantum-classical wavelength very important for obtaining high performance 
QN operations.

\subsection{Quantum State Tomography of Entangled Photon-Classical Coexistence}

\begin{figure}[tp]
\centering
\includegraphics[width=0.45\textwidth]{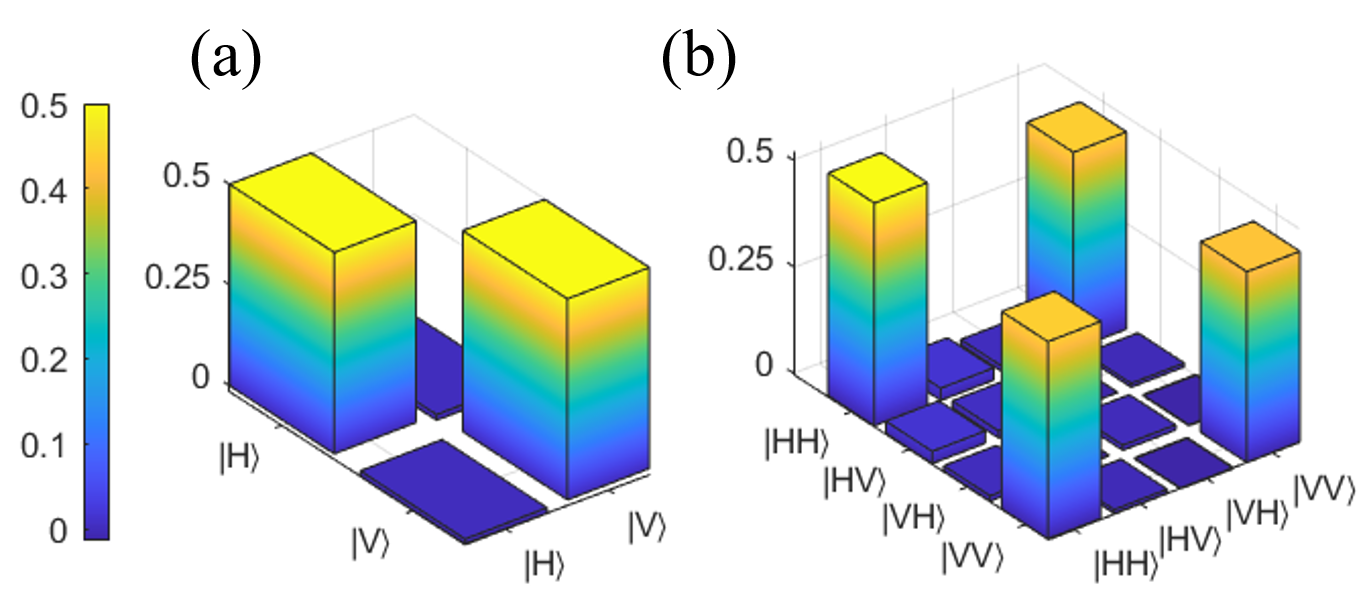}
\caption{(a) Single qubit QST of spRS noise from the WDM C-band source. (b) QST of polarization entangled photon pair-classical coexistence when the 1287 nm channel co-propagates 47.9 km installed fiber with 18.1 dBm WDM C-band launch power.}
\label{fig4}
\end{figure}

To characterize the effect of spRS noise on the polarization entangled state, we perform 36 projective measurement QST between a $1287 \mathrm{~nm}$ photon transmitting $47.9 \mathrm{~km}$ and $1313 \mathrm{~nm}$ routed over $5.4 \mathrm{~km}$ spooled fiber at a CCR of $30.1 \ \mathrm{ccps}$. We first measure single qubit tomography on the spRS noise alone (Fig. \ref{fig4}(a)), where we obtain a purity $P=\operatorname{tr}\left(\rho^2\right)=0.50$, indicating that the spRS noise is effectively a completely mixed quantum state $(P=0.5)$. Therefore, spRS from the WDM C-band source appears to be polarization independent in the co-propagating channel. Although Raman gain is slightly polarization dependent at O-band/C-band offsets \cite{stolen}, this is not unexpected due to polarization mode dispersion (particularly in WDM classical systems) and is in line with recent quantum channel descriptions of spRS in single-frequency C-band/C-band coexistence \cite{PhysRevApplied.19.044026}. Performing such studies on O-band/C-, L-band coexistence and WDM CC coexistence generally would expand this channel description to more quantum-classical scenarios.\\ \indent We then characterize the impact of spRS noise on the EPPs, where we perform 2-qubit QST on the entangled photons. We obtain a fidelity F$\left(\Phi^{+}, \rho_{\text {dark}}\right)=97.7 \%$ to the $\left|\Phi^{+}\right\rangle$ Bell state in dark fiber, negligibly different from back-to-back QST without accidental count subtraction. Next, we copropagate the C-band source and measure the EPP-classical coexistence density matrix $\rho_{\text {coex}}$ for each power setting. Fig. \ref{fig4}(b) shows the real part of $\rho_{\text {coex}}$ when co-propagating $\mathrm{P}_{\mathrm{Tx}}$ $=18.1 \ \mathrm{dBm}$. To evaluate the impact of spRS, we calculate the fidelity of $\rho_{\text {coex}}$ to $\rho_{\text {dark}}$, where we obtain $\mathrm{F}_{\text {coex}}\left(\rho_{\text {dark}}, \rho_{\text {coex}}\right)=$ $99.6 \%, 99.1 \%$, and $98.5 \%$ with $\mathrm{P}_{\mathrm{Tx}}=14,16.2$, and $18.1 \ \mathrm{dBm}$, corresponding to only a $0.4 \%, 0.9 \%$, and $1.5 \%$ reduction, respectively.

\section{Discussion}

\subsection{Further Improvements}\label{3a}

Although we achieved relatively high visibility across the full 1282 nm-1318 nm region, we clearly demonstrated that some quantum-classical channels are less impacted by spRS than others. In terms of coexisting with WDM CCs, it is important to consider the ability to reduce noise in the context of how this restricts the number of channels in each system. Even though the <1300 nm quantum channels were optimized with a broad BW $(\sim 16 \mathrm{~nm})$ WDM C-band system, Fig. \ref{fig:fig1}(a) indicates that the $\sim 1282 \mathrm{~nm}$ quantum channels would have performed similarly to this experiment co-propagating alongside a fully packed $1530 \mathrm{~nm}-1565 \mathrm{~nm}$ WDM CC system, although the other $<1300 \mathrm{~nm}$ quantum channels would be slightly more impacted from the $<1550 \mathrm{~nm}$ classical channels. Alternatively for all cases, the classical BW can always be increased by $60 \mathrm{~nm}$ using L-band CCs with little additional noise impact.

Beyond wavelength engineering, there are many modifications that could significantly improve the results here. The tunable filters used to characterize EPPs came at the expense of $6 \mathrm{~dB}$ total loss. Using lower loss $(\sim 0.5 \mathrm{~dB})$ filters would improve quantum rates by a factor of $\sim 3$ for a particular $\mu_{\mathrm{c}}$, increase the maximum visibility, and reduce multi-photon generation for the CCRs achieved in this study.

Even more importantly for the engineering of future QNs, narrower time-frequency filtering to a single spectral-temporal mode and gated detection would dramatically improve spRS noise resilience. These improvements will be naturally found in advanced QNs as single mode identical photons are needed for high visibility Hong-Ou-Mandel interference in BSMs. Further, single mode EPPs do not suffer from multi-mode multi-pair emission, allowing higher visibilities even in dark fiber \cite{takesue_effects_2010}. Using a $\sim 100 \ \mathrm{ps}$ coincidence window and $5 \ \mathrm{GHz}$ filters would reduce spRS rates by $\sim 60$ times compared to this work. This will allow many 100 's of $\mathrm{mW}$ co-propagating power throughout the full C-band to have little impact on $<1310 \mathrm{~nm}$ quantum signals, thus posing little impediment to co-existence of quantum channels with typical fully populated EDFA-amplified classical data streams.

\subsection{Implications for Future Quantum Networking}

Although we demonstrated the transmission of one of the entangled photons over the lit fiber, there can be situations in which both photons are transmitted over the same or different lit fiber with each either co-, counter-, or bi-directionally propagating with the same or different classical signals. Further, each may have different signal, idler, and classical wavelength offsets, where different QN configurations and routing will be impacted by spRS more than others. This leads to an interesting analysis of the optimal ways to design spRS robust physical layers and optimized routing of photons in switch-based QNs \cite{alshowkan_reconfigurable_2021, chung_design_2022}. Note that since Raman gain is non-uniform across all telecom bands, such considerations are applicable to all frequency offsets including C-band/C-band coexistence, where the concepts described in this paper can be similarly applied.

Beyond EPP distribution, quantum-classical coexistence becomes even more complex for quantum teleportation, entanglement swapping, repeaters, and beyond. For example, swapping involves 4 photons transmitted over 4 independent fibers where each could have coexisting classical channels, and chains of repeaters could reach arbitrary numbers of fiber links and coexisting channels. The rate of $n$-fold accidental coincidence detection is proportional to the product of the singles counts in each channel $\left(\mathrm{A}_{\text{n-fold}}\right.$ $\propto \mathrm{S}_1 \ldots \mathrm{S}_n$ \cite{PhysRev.53.752}). For $k$ of $n$ photons in a system coexisting with CCs, reducing spRS rates by a factor of $m$ in each channel leads to terms in the expansion of $\mathrm{A}_{\mathrm{n} \text {-fold }}$ involving   $k$ spRS photons being reduced by $m^{\mathrm{k}}$. This is the case for reducing filter BWs, temporal coincidence windows, or comparing the spRS rates based on the choice of quantum-classical wavelengths. Considering the difference in spRS when choosing the optimal quantum-classical wavelengths, the probability of accidental coincidences involving $k$ spRS photons is reduced by $k$ orders of magnitude. We expect such ideas will be extremely valuable in general and that a broad research field will arise to evaluate the ways in which spRS impacts future QN physical layers and how to optimize QNs in the control plane.

Lastly, we note that the optimal frequency offsets in anti-Stokes spRS arise in Stokes spRS too. i.e., C- or L-band quantum states are more robust to O-band classical light at the reversed offsets explored here \cite{thomas_ofc_2023, Burenkov:23}. This can be used in C-/L-band QNs for classical light-based time synchronization \cite{Burenkov:23,kapoor_picosecond_2023}, communication of quantum measurement results between nodes, or integrating O-band CCs to ease the cost of C- or L-band QNs. However, since anti-Stokes scattering has fundamentally less noise than Stokes scattering, the wavelength combinations explored in this paper represent the lowest spRS noise that can be achieved when wavelength engineering telecom band quantum-classical systems.

\section{Conclusion}

We have demonstrated for the first time the ability to achieve high-fidelity O-band polarization entangled photon pair distribution coexisting in 47.9 km of installed standard fiber with at least 18 dBm wavelength division multiplexed C-band power, on the order required for terabits per second classical data rates and more than an order of magnitude higher power than demonstrated to date with entangled photons. We characterized multiple 50 GHz channel pairs spanning 1282 nm-1318 nm and provided a framework for noise optimizing quantum and classical wavelengths, quantum network routing, and measurement systems to allow multi-channel O-band quantum and C-, L-band classical networking in the same fibers with the lowest spRS noise impact. We further characterized the interplay between spRS and multi-photon pair emission noise in entangled photon pair-based quantum applications as well as quantum state tomography of the spRS noise and polarization entangled photons.

In total, these results demonstrate the promise of O-band quantum systems for deploying advanced quantum network capabilities beyond dark fibers, even in the more extreme cases such as in fibers constituting the classical backbone infrastructure. Further, we believe our analysis of the spRS at various wavelength offsets will have wide appeal in the field of quantum-classical coexistence of arbitrary quantum states, and likely critical for quantum functions even more susceptible to added noise such as quantum teleportation, entanglement swapping, etc., paving the way for fulfilling the full potential of both quantum and classical internets coexisting in the same fiber infrastructure.

\begin{backmatter}
\bmsection{Funding} This work is funded by Subcontract No. 664603 from Fermi Research Alliance, LLC (FRA) to Northwestern University issued under Prime Contract No. DE-AC02-07CH11359 between FRA and the U.S. Department of Energy (DOE). Although the work is supported by the DOE’s Advanced Scientific Computing Research Transparent Optical Quantum Networks for Distributed Science program, no government endorsement is implied.

\bmsection{Acknowledgments} The authors would like to thank the Illinois Express Quantum Network (IEQNET) team, which is a collaboration between Northwestern University, Fermilab National Laboratory, Argonne National Laboratory, Caltech, and NuCrypt LLC to deploy a quantum network within the Chicago, IL, USA metropolitan area.

\bmsection{Disclosures} The authors declare no conflicts of interest.

\bmsection{Data Availability Statement} 

Data may be obtained from the authors upon reasonable request. 

\end{backmatter}

\bibliography{refOPTICAPAPER}




\end{document}